\documentstyle[12pt,epsfig]{article}
\begin{document}
\setlength{\parskip}{0.45cm}
\setlength{\baselineskip}{0.75cm}
%XXXXXXXXXXXXXXXXXXXXXXXXXXXXXXXXXXXXXX
%
%SETTINGS FOR PREPRINT-SPACED VERSION
%setlength{\parskip}{0.45cm}
%setlength{\baselineskip}{0.75cm}
%
% SETTINGS FOR DOUBLE - SPACED VERSION
%\setlength{\parskip}{0.65cm}
%\setlength{\baselineskip}{0.95cm}
%
%XXXXXXXXXXXXXXXXXXXXXXXXXXXXXXXXXXXXXX
\begin{titlepage}
\setlength{\parskip}{0.25cm}
\setlength{\baselineskip}{0.25cm}
\begin{flushright}
DO-TH 99/01\\  
%\vspace{0.2cm}
%hep--ph/9902...\\
\vspace{0.2cm}
January 1999
\end{flushright}
\vspace{1.0cm}
\begin{center}
\LARGE
{\bf Pionic Parton Distributions Revisited}
\vspace{1.5cm}

\large
M. Gl\"uck, E.\ Reya and I.\ Schienbein\\
\vspace{1.0cm}

\normalsize
{\it Institut f\"{u}r Physik, Universit\"{a}t Dortmund}\\ 
{\it D-44221 Dortmund, Germany} \\

\vspace{1.5cm}
\end{center}
\begin{abstract}
Using constituent quark model constraints we calculate the gluon and
sea--quark content of pions solely in terms of their valence density
(fixed by $\pi N$ Drell--Yan data) and the known sea and gluon distributions
of the nucleon, using the most recent updated valence--like input parton
densities of the nucleon.  The resulting small--$x$ dynamical QCD 
predictions for $g^{\pi}(x,Q^2)$ and $\bar{q}\,^{\pi}(x,Q^2)$ are unique
and parameter free.  Simple analytic parametrizations of the resulting
parton distributions of the pion are presented in LO and NLO.  These
results and parametrizations will be important, among other things,
for updated formulations of the parton distributions of real and virtual
photons.
\end{abstract}
\end{titlepage}
%
%\section{Introduction}
The parton content of the pion is poorly known at present.  The main
experimental source about these distributions is mainly due to data of
Drell--Yan dilepton production in $\pi^-$--tungsten reactions 
\cite{ref1,ref2,ref3}, which determine the shape of the pionic valence
density $v^{\pi}(x,Q^2)$ rather well, and due to measurements of direct
photon production in $\pi^{\pm}p\to\gamma X$ \cite{ref1,ref4} which
constrain the pionic gluon distribution $g^{\pi}(x,Q^2)$ only in the
large--$x$ region \cite{ref5}.  
In general, however, present data are not sufficient for fixing $g^{\pi}$
uniquely, in particular the pionic sea density $\bar{q}\,^{\pi}(x,Q^2)$
remains entirely unconstrained experimentally.  
Therefore we have previously \cite{ref6} utilized a constituent quark 
model \cite{ref7} to relate $\bar{q}\,^{\pi}$ and $g^{\pi}$ to the much 
better known radiatively generated parton distributions $f^p(x,Q^2)$ of 
the proton \cite{ref8}.
These relations arise as follows:  describing the constituent quark
structure of the proton $p=UUD$ and the pion, say $\pi^+=U\bar{D}$,
by the scale $(Q^2)$ independent distributions $U^{p,\pi^+}(x),\,\,
D^p(x)$ and $\bar{D}\,^{\pi^+}(x)$, and their universal (i.e.\ hadron 
independent) partonic content by $v_c(x,Q^2),\,\, g_c(x,Q^2)$ and 
$\bar{q}_c(x,Q^2)$, the usual parton content of the proton and the pion 
is then given by
%Eqs.(1) und (2)
\begin{eqnarray}
f^p(x,Q^2) & = & \int_x^1\frac{dy}{y}\left[ U^p(y)+D^p(y)\right] f_c
 \left( \frac{x}{y}, Q^2\right)\\
f^{\pi}(x,Q^2) & = & \int_x^1\frac{dy}{y} \left[ U^{\pi^+}(y)+
  \bar{D}\,^{\pi^+}(y)\right] f_c \left( \frac{x}{y}, Q^2\right)
\end{eqnarray}
where $f=v,\,\bar{q},\,g$ with $v^p=u_v^p+d_v^p,\,\,\, 
\bar{q}\,^p=(\bar{u}\,^p+\bar{d}\,^p)/2,\,\,\, 
v^{\pi}=u_v^{\pi^+}+\bar{d}\,_v^{\pi^+},\,\,
\bar{q}\,^{\pi}=(\bar{u}\,^{\pi^+}+d^{\pi^+})/2$ and $\bar{u}\,^{\pi^+}
=d^{\pi^+}$ due to ignoring minor SU(2)$_{\rm{flavor}}$ breaking effects
in the pion `sea' distributions. Assuming these relations to apply at
the low resolution scale $Q^2=\mu^2$ ($\mu^2_{\rm{LO}} = 0.23$ GeV$^2,\,\,\,
\mu^2_{\rm{NLO}}=0.34$ GeV$^2$) of \cite{ref8} where the strange quark
content was considered to be negligible,  
%Eq.(3)
\begin{equation}
 s^p(x,\mu^2) = \bar{s}\,^p(x,\mu^2) =s^{\pi}(x,\mu^2) =
  \bar{s}\,^{\pi}(x,\mu^2) = 0,
\end{equation}
one obtains from (1) and (2) the constituent quark independent
relations \cite{ref6}  
%Eq.(4)
\begin{equation}
\frac{v^{\pi}(n,\mu^2)}{v^p(n,\mu^2)} =
\frac{\bar{q}\,^{\pi}(n,\mu^2)}{\bar{q}\,^p(n,\mu^2)} =
\frac{g^{\pi}(n,\mu^2)}{g^p(n,\mu^2)}
\end{equation}
where for convenience we have taken the Mellin $n$-moments of eqs.\ (1) 
and (2), i.e.\ $f(n,Q^2)\equiv \int_0^1x^{n-1}f(x,Q^2)dx$.  Thus, as
soon as $v^{\pi}(x,\mu^2)$ is reasonably well \mbox{determined} from 
experiment,
our basic relations (4) uniquely fix the gluon and sea densities of the
pion in terms of the rather well known parton distributions of the 
proton:
%Eq.(5)
\begin{equation}
g^{\pi}(n,\mu^2) = \frac{v^{\pi}(n,\mu^2)}{v^p(n,\mu^2)}\, g^p(n,\mu^2),
\quad\quad
\bar{q}\,^{\pi}(n,\mu^2) = \frac{v^{\pi}(n,\mu^2)}{v^p(n,\mu^2)}\,
\bar{q}\,^p(n,\mu^2).
\end{equation}

Furthermore, the sum rules \cite{ref6}
%Eqs.(6) and (7)
\begin{eqnarray}
\int_0^1 v^{\pi}(x,Q^2) dx & = & 2\\
\int_0^1 xv^{\pi}(x,Q^2)dx & = & \int_0^1xv^p(x,Q^2)dx
\end{eqnarray}
impose strong constraints on $v^{\pi}(x,\mu^2)$ which are very useful
for its almost unambiguous determination from the $\pi N$ Drell--Yan
data.  Notice that eq.\ (7), together with (4), implies the 
energy--momentum sum rule for $f^{\pi}$ to be manifestly satisfied.
In addition, eq.\ (7) implies that the valence quarks in the proton
and the pion carry similar total fractional momentum as suggested
by independent analyses within the framework of the radiative parton
model \cite{ref5,ref8}.

The relations in eq.\ (5) imply that any updating of $f^p(x,\mu^2)$
yields a corresponding updating of $f^{\pi}(x,\mu^2)$.  Recently an
updating of $f^p(x,\mu^2)$ within the framework of the radiative
(dynamical) parton model was undertaken \cite{ref9} utilizing additional
improved data on $F^p_2(x,Q^2)$ from HERA \cite{ref10,ref11} and a somewhat
increased $\alpha_s(M_Z^2) = 0.114$ resulting in a slight increase in
$\mu^2\,\, (\mu^2_{\rm{LO}}=0.26$ GeV$^2,\,\,
\mu^2_{\rm{NLO}} = 0.40$ GeV$^2$).  An improved treatment of the
running $\alpha_s(Q^2)$ at low $Q^2$ was furthermore implemented by
solving in NLO$(\overline{\rm{MS}})$
%Eq.(8)
\begin{equation}
\frac{d\alpha_s(Q^2)}{d\, ln\, Q^2} = -\frac{\beta_0}{4\pi}\,
  \alpha_s^2(Q^2)-\frac{\beta_1}{16\pi^2}\,\alpha_s^3(Q^2)
\end{equation}
numerically \cite{ref9} rather than using the approximate NLO solution
%Eq.(9)
\begin{equation}
\frac{\alpha_s(Q^2)}{4\pi} \simeq \frac{1}{\beta_0\, ln\, (Q^2/\Lambda^2)}
 \, - \,\, \frac{\beta_1}{\beta_0^3}\,\,
   \frac{ln\, ln\, (Q^2/\Lambda^2)}{ln^2\, (Q^2/\Lambda^2)}
\end{equation}
as done in \cite{ref5,ref6,ref8}, which is sufficiently accurate only
for $Q^2$ \raisebox{-0.1cm}{$\stackrel{>}{\sim}$} $m_c^2\simeq2$ GeV$^2$ 
\cite{ref9}.  The LO and
NLO evolutions of $f^{\pi}(n,Q^2)$ to $Q^2>\mu^2$ are performed in
Mellin $n$--moment space, followed by a straightforward numerical
Mellin--inversion \cite{ref12} to Bjorken-x space.  It should be noted
that the evolutions are always performed in the fixed (light) $f=3$
flavor factorization scheme \cite{ref13,ref6,ref8,ref9}, 
i.e.\ we refrain from generating radiatively massless `heavy' quark
densities $h^{\pi}(x,Q^2)$ where $h=c,\, b$, etc., in contrast to
\cite{ref5}.  Hence heavy quark contributions have to be calculated
in fixed--order perturbation theory via, e.g., $g^{\pi}g^p\to h\bar{h},\,\,
\bar{u}\,^{\pi}u^p\to h\bar{h}$, etc.  (Nevertheless, rough estimates
of `heavy' quark effects, valid to within a factor of 2, say, can be
easier obtained with the help of the massless densities $c^{\pi}(x,Q^2)$
and $b^{\pi}(x,Q^2)$ given in \cite{ref5}.)

Using all these modified ingredients together with the new updated 
\cite{ref9} $f^p(x,\mu^2)$ in our basic predictions in eq.\ (5), the
present reanalysis of the available Drell--Yan data \cite{ref2},
closely following the procedure described in \cite{ref6}, yields
%Eqs.(10) and (11)
\begin{eqnarray}
v_{\rm{LO}}^{\pi}(x,\mu_{\rm{LO}}^2) & = & 1.129x^{-0.496}(1-x)^{0.349}
   (1+0.153\sqrt{x})\\
v_{\rm{NLO}}^{\pi}(x,\mu_{\rm{NLO}}^2) & = & 1.391x^{-0.447}(1-x)^{0.426}
\end{eqnarray}
where 
\cite{ref9} $\mu_{\rm{LO}}^2=0.26$ GeV$^2$ and $\mu_{\rm{NLO}}^2=0.40$
GeV$^2$.  These updated input valence densities correspond to total
momentum fractions
%Eqs.(12) and (13)
\begin{eqnarray}
\int_0^1x\, v_{\rm{LO}}^{\pi}(x,\mu_{\rm{LO}}^2)dx & = & 0.563\\
\int_0^1x\, v_{\rm{NLO}}^{\pi}(x,\mu_{\rm{NLO}}^2)dx & = & 0.559
\end{eqnarray}
as dictated by the valence densities of the proton \cite{ref9} via eq.\ (7).
Our new updated input distributions in eqs.\ (10), (11) and (5) are rather
different than the original GRV$_{\pi}$ input \cite{ref5} in fig.\ 1 which
is mainly due to the vanishing sea input of GRV$_{\pi}$ in contrast to the
present one in eq.\ (5).  On the other hand, our updated input in fig.\ 1
is, as expected, rather similar to the one of \cite{ref6}.  In both cases,
however, the valence and gluon distributions become practically 
indistinguishable from our present updated ones at scales relevant for
present Drell--Yan dimuon and direct--$\gamma$ production data, 
$Q^2\equiv M_{\mu^+\mu^-}^2\simeq20$ GeV$^2$, as illustrated in fig.\ 2.
Therefore our present updated pionic distributions give an equally good
description of all available $\pi N$ Drell--Yan data as the ones shown
in \cite{ref6}.

For completeness let us mention that our basic predictions (5) 
for the valence--like gluon and sea densities at $Q^2=\mu^2$, as shown 
in fig.\ 1, can be simply parametrized in Bjorken--$x$ space :
in LO at $Q^2=\mu_{\rm{LO}}^2=0.26$ GeV$^2$
%eq. (14)
\begin{eqnarray}
x\, g^{\pi}(x,\mu_{\rm{LO}}^2) & = & 7.326\,x^{1.433} (1-1.919\, 
       \sqrt{x}+1.524\,x)(1-x)^{1.326}\nonumber\\
x\, \bar{q}\,^{\pi}(x,\mu_{\rm{LO}}^2) & = &  0.522\,x^{0.160} 
   (1- 3.243\, \sqrt{x} + 5.206\, x) (1-x)^{5.20}\, ,
\end{eqnarray}
whereas in NLO at $Q^2=\mu_{\rm{NLO}}^2 = 0.40$ GeV$^2$ we get
%eq. (15)
\begin{eqnarray}
x\, g^{\pi}(x,\mu_{\rm{NLO}}^2) & = & 5.90\, x^{1.270}(1-2.074\,\sqrt{x}\,
     +1.824\, x)(1-x)^{1.290}\nonumber\\
x\, \bar{q}\,^{\pi}(x,\mu_{\rm{NLO}}^2) & = & 0.417\, x^{0.207}(1-2.466\,
     \sqrt{x}\, +3.855\, x)(1-x)^{4.454} .
\end{eqnarray}

Finally, fig.\ 3 shows our resulting predictions for $x\, g^{\pi}(x,Q^2)$
and $x\,\bar{q}\,^{\pi}(x,Q^2)$ as compared to the former GRV$_{\pi}$ 
results \cite{ref5}.  The GRV$_{\pi}$ results for $x\, \bar{q}\,^{\pi}$
are significantly steeper and softer for $x$ \raisebox{-0.1cm}{$\stackrel{>}
{\sim}$} 0.01
due to the vanishing SU(3)$_{\rm{flavor}}$ symmetric (light) sea input
$x\,\bar{q}\,^{\pi}(x,\mu^2)=0$, in contrast to our present approach
\cite{ref6} based on a more realistic finite light sea input in eq.\ (5).
The valence--like gluon and sea inputs at $Q^2=\mu^2$, which become
(vanishingly) small at $x<10^{-2}$, are also shown in fig.\ 3.  This
illustrates again the purely dynamical origin of the small--$x$ structure
of gluon and sea quark densities at $Q^2>\mu^2$.  Our predictions for
$s^{\pi}=\bar{s}\,^{\pi}$, as evolved from the vanishing input in eq.\ (3),
are not shown in the figure since they practically coincide with
$\bar{q}\,^{\pi}(x,Q^2)$ of GRV$_{\pi}$ shown in fig.\ 3 which also results
from a vanishing input \cite{ref5}.  Simple analytic parametrizations
of our LO and NLO predictions for $f^{\pi}(x,Q^2)$ are given in the Appendix.

To conclude let us recall that an improvement of $f^{\pi}(x,Q^2)$ is
particularly important in view of its central role in the construction
of the photon structure function and the photonic parton distributions
\cite{ref14,ref15,ref16,ref17,ref18}.  Furthermore, recent (large rapidity
gap) measurements of leading proton and neutron production in deep inelastic
scattering at HERA \cite{ref19} allow, under certain (diffractive) model
assumptions, to constrain and test the pion structure functions for the
first time at far smaller vales of $x$ (down to about $10^{-3}$) than
those attained from fixed target $\pi N$ experiments.
\newpage

\noindent{\Large{\bf{Acknowledgement}}}

\noindent This work has been supported in part by the `
Bundesministerium f\"ur
Bildung, Wissenschaft, Forschung und Technologie', Bonn.
\vspace{1.5cm}

\noindent{\Large{\bf{Appendix}}}
\renewcommand{\theequation}{A.\arabic{equation}}
\setcounter{equation}{0}

\noindent {\bf{A.}}  Parametrization of LO parton distributions\\
Defining \cite{ref9}
%Eq.(A.1)
\begin{equation}
s\equiv ln\, \frac{ln\,[Q^2/(0.204\, {\rm{GeV}})^2]}
    {ln\,[\mu_{\rm{LO}}^2/(0.204\, {\rm{GeV}})^2]}
\end{equation}

\noindent to be evaluated for $\mu_{\rm{LO}}^2=0.26$ GeV$^2$, all our 
resulting
pionic parton distributions can be expressed by the following simple
parametrizations,
valid for $0.5$ \raisebox{-0.1cm}{$\stackrel{<}{\sim}$} $Q^2$ 
\raisebox{-0.1cm} {$\stackrel{<}{\sim}$} $10^5$ GeV$^2$ 
\mbox{(i.e. $0.31 \leq s$ \raisebox{-0.1cm}{$\stackrel{<}{\sim}$} 2.2)} 
and $10^{-5}$ \raisebox{-0.1cm}{$\stackrel{<}{\sim}$} $x\,<\, 1$.  
For the valence distribution we take
%Eq.(A.2)
\begin{equation}
x\, v^{\pi}(x,Q^2) = N\, x^a(1+A\sqrt{x}+Bx)(1-x)^D
\end{equation}
with
%Eq.(A.3)
\begin{eqnarray}
N & = & 1.212 + 0.498\,s + 0.009\, s^2 \nonumber\\
a & = & 0.517 - 0.020\,s \nonumber\\
A & = & -0.037 - 0.578\,s\nonumber\\
B & = & 0.241 + 0.251\,s\nonumber\\
D & = & 0.383 + 0.624\,s\, .
\end{eqnarray}
The gluon and light sea--quark distributions are parametrized as
%Eq.(A.4)
\begin{equation}
x\, w^{\pi}(x,Q^2)=\left[ x^a \left( A+B\sqrt{x} +Cx \right) 
   \left( ln\, \frac{1}{x} \right)^b + s^{\alpha} \,\,{\rm{exp}}
     \left( -E+\sqrt{E's^{\beta}ln\, \frac{1}{x}} \right)\right]
       (1-x)^D.
\end{equation}
\pagebreak

For $w=g$
%Eq.(A.5)
\begin{equation}
\begin{array}{lcllcl}
\alpha & = & 0.504, & \beta & = & 0.226,\\[1mm]
a & = & 2.251 - 1.339\,\sqrt{s}, &  b & = & 0,\\[1mm]
A & = & 2.668 - 1.265\, s + 0.156\, s^2, & B  & = & -1.839 + 0.386
  \,s,\\[1mm]
C & = & -1.014 + 0.920\,s - 0.101\, s^2, &  D & = & -0.077 + 1.466
  \,s,\\[1mm]
E & = & 1.245 + 1.833\,s, & E' &  = & 0.510 + 3.844\,s\,,
\end{array}
\end{equation}
and for the light sea $w=\bar{q}$
%Eq.(A.6)
\begin{equation}
\begin{array}{lcllcl}
\alpha & = & 1.147,  & \beta & = & 1.241,\\[1mm]
a & = & 0.309 - 0.134\,\sqrt{s}, &  b & = & 0.893-0.264\, \sqrt{s},\\[1mm]
A & = & 0.219 - 0.054\, s, &  B & = & -0.593 + 0.240\,s,\\[1mm]
C & = & 1.100 - 0.452\,s, & D & = & 3.526 + 0.491\,s,\\[1mm]
E & = & 4.521 + 1.583\, s, &  E' & = & 3.102\, .
\end{array}
\end{equation}
The strange sea distribution $s^{\pi}=\bar{s}\,^{\pi}$ is parametrized 
as
%Eq.(A.7)
\begin{equation}
x\bar{s}\,^{\pi}(x,Q^2) = \frac{s^{\alpha}}{(ln\,\frac{1}{x})^a}\,
\left( 1 + A\sqrt{x}+Bx\right)(1-x)^D\, {\rm{exp}}\, 
\left( -E+\sqrt{E's^{\beta}ln\frac{1}{x}}\, \right)
\end{equation}
with
%Eq.(A.8)
\begin{equation}
\begin{array}{lcllcl}
\alpha & = & 0.823,  & \beta & = & 0.650,\\[1mm]
a & = & 1.036 - 0.709\,s, &  A & = & -1.245+0.713\,s,\\[1mm]
B & = & 5.580 - 1.281\,s, &  D & = & 2.746 - 0.191\,s,\\[1mm]
E & = & 5.101 + 1.294\,s, &  E' & = & 4.854-0.437\, s\, .
\end{array}
\end{equation}
\vspace{0.5cm}

\noindent {\bf{B.}}  Parametrization of NLO($\overline{{\rm{MS}}}$)
parton distributions\\
Defining \cite{ref9}
%Eq.{A.9}
\begin{equation}
s\equiv ln\, \frac{ln\,[Q^2/(0.299\, {\rm{GeV}})^2]}
      {ln\, [\mu_{\rm{NLO}}^2/(0.299\, {\rm{GeV}})^2]}
\end{equation}

\noindent to be evaluated for $\mu_{\rm{NLO}}^2=0.40$ GeV$^2$, our NLO 
predictions
can be parametrized as the LO ones and are similarly valid for
0.5 \raisebox{-0.1cm}{$\stackrel{<}{\sim}$} $Q^2$ 
\raisebox{-0.1cm}{$\stackrel{<}{\sim}$} $10^5$ GeV$^2$ 
(i.e. 0.14 \raisebox{-0.1cm}{$\stackrel{<}{\sim}$} s 
\raisebox{-0.1cm}{$\stackrel{<}{\sim}$} 2.38) 
and \mbox{$10^{-5}$ \raisebox{-0.1cm}{$\stackrel{<}{\sim}$} $x\,<\, 1$}.
The valence distribution is given by (A.2) with
%Eq.(A.10)
\begin{eqnarray}
N & = & 1.500 + 0.525\,s - 0.050\, s^2 \nonumber\\
a & = & 0.560 - 0.034\,s \nonumber\\
A & = & -0.357 - 0.458\,s\nonumber\\
B & = & 0.427 + 0.220\,s\nonumber\\
D & = & 0.475 + 0.550\,s\, .
\end{eqnarray}
The gluon and light sea distributions are parametrized as in (A.4) 
where for $w=g$
%Eq.(A.11)
\begin{equation}
\begin{array}{lcllcl}
\alpha & = & 0.793, & \beta & = & 1.722,\\[1mm]
a & = & 1.418 - 0.215\sqrt{s}, &  b & = & 0,\\[1mm]
A & = & 5.392 + 0.553\,s - 0.385\, s^2, & B  & = & -11.928 + 1.844\,s,\\[1mm]
C & = & 11.548 - 4.316\,s + 0.382\, s^2, &  D & = & 1.347 + 1.135\,s,\\[1mm]
E & = & 0.104 + 1.980\,s, & E' &  = &2.375 - 0.188\,s\,.
\end{array}
\end{equation}
and for the light sea $w=\bar{q}$
%Eq.(A.12)
\begin{equation}
\begin{array}{lcllcl}
\alpha & = & 1.118,  & \beta & = & 0.457,\\[1mm]
a & = & 0.111 - 0.326\,\sqrt{s}, &  b & = & -0.978-0.488\,\sqrt{s},\\[1mm]
A & = & 1.035 - 0.295\,s, & B & = & -3.008 + 1.165 \,s,\\[1mm]
C & = & 4.111 - 1.575\,s, &  D & = & 6.192 + 0.705\,s,\\[1mm]
E & = & 5.035 + 0.997\,s, &  E' & = & 1.486 + 1.288\,s\, .
\end{array}
\end{equation}
The strange sea distribution is parametrized as in (A.7) with
%Eq.(A.13)
\begin{equation}
\begin{array}{lcllcl}
\alpha & = & 0.908,  & \beta & = & 0.812,\\[1mm]
a & = & -0.567 - 0.466\,s, &  A & = & -2.348 + 1.433\,s,\\[1mm]
B & = & 4.403, & D & = & 2.061,\\[1mm]
E & = & 3.796 + 1.618\,s, &  E' & = & 0.309 + 0.355\,s\, .
\end{array}
\end{equation}

Let us recall that in the light quark sector $u_v^{\pi^+}=\bar{d}\,_v^{\pi^+}
= \bar{u}\,_v^{\pi^-}=d_v^{\pi^-},\,\,\,
\bar{u}\,^{\pi^+}=d^{\pi^+} = u^{\pi^-}=\bar{d}\,^{\pi^-}$ and
$f^{\pi^0}=(f^{\pi^+}+f^{\pi^-})/2$.

\newpage

\noindent{\Large{\bf{\underline{Figure Captions}}}}
\begin{itemize}
\item[\bf{Fig.\ 1}]  The valence and valence--like input distributions
	$xf^{\pi}(x,Q^2=\mu^2)$ with $f=v,\, \bar{q},\,g$ as compared 
	to those of GRV$_{\pi}$ \cite{ref5}.  Notice that GRV$_{\pi}$
	employs a vanishing SU(3)$_{\rm{flavor}}$ symmetric $\bar{q}\,^{\pi}$
	input at $\mu_{\rm{LO}}^2=0.25$ GeV$^2$ and 
	$\mu_{\rm{NLO}}^2=0.3$ GeV$^2$ \cite{ref5}.  Our present 
	SU(3)$_{\rm{flavor}}$ broken sea densities refer to a vanishing 
	$s^{\pi}$ input in (3), as for GRV$_{\pi}$ \cite{ref5}.  
\item[\bf{Fig.\ 2}]  Comparison of our NLO valence distribution at 
	$Q^2=20$ GeV$^2$ with the one of GRV$_{\pi}$ \cite{ref5} and
	GRS \cite{ref6}.  This density plays the dominant role for 
	describing presently available $\pi N$ Drell--Yan dimuon production
	data.  For illustration, the gluon and sea densities are shown as
	well.  The SU(3)$_{\rm{flavor}}$ symmetric GRV$_{\pi}$ sea
	$\bar{q}\,^{\pi}=s^{\pi}$ is not shown, since it is similar to
	$s^{\pi}$ of our present analysis and of GRS which are all
	generated from a vanishing input at $Q^2=\mu^2$, cf.\ eq.\ (3).
\item[\bf{Fig.\ 3}]  The small--$x$ predictions of our radiatively
	generated pionic gluon and sea--quark distributions in LO and NLO
	at various fixed values of $Q^2$ as compared to those of 
	GRV$_{\pi}$ \cite{ref5}.  The valence--like inputs, according to
	eq.\ (5) as presented in fig.\ 1, are shown for illustration by
	the lowest curves referring to $\mu^2$.  The predictions for
	the strange sea density $s^{\pi}=\bar{s}\,^{\pi}$ are similar to
	the GRV$_{\pi}$ results for $\bar{q}\,^{\pi}$.  The results are
	multiplied by the numbers indicated in brackets.    
\end{itemize}

\newpage
\pagestyle{empty}
\begin{figure}
\centering
%\vspace*{-1cm}
%\hspace*{-1.5cm}
\epsfig{figure=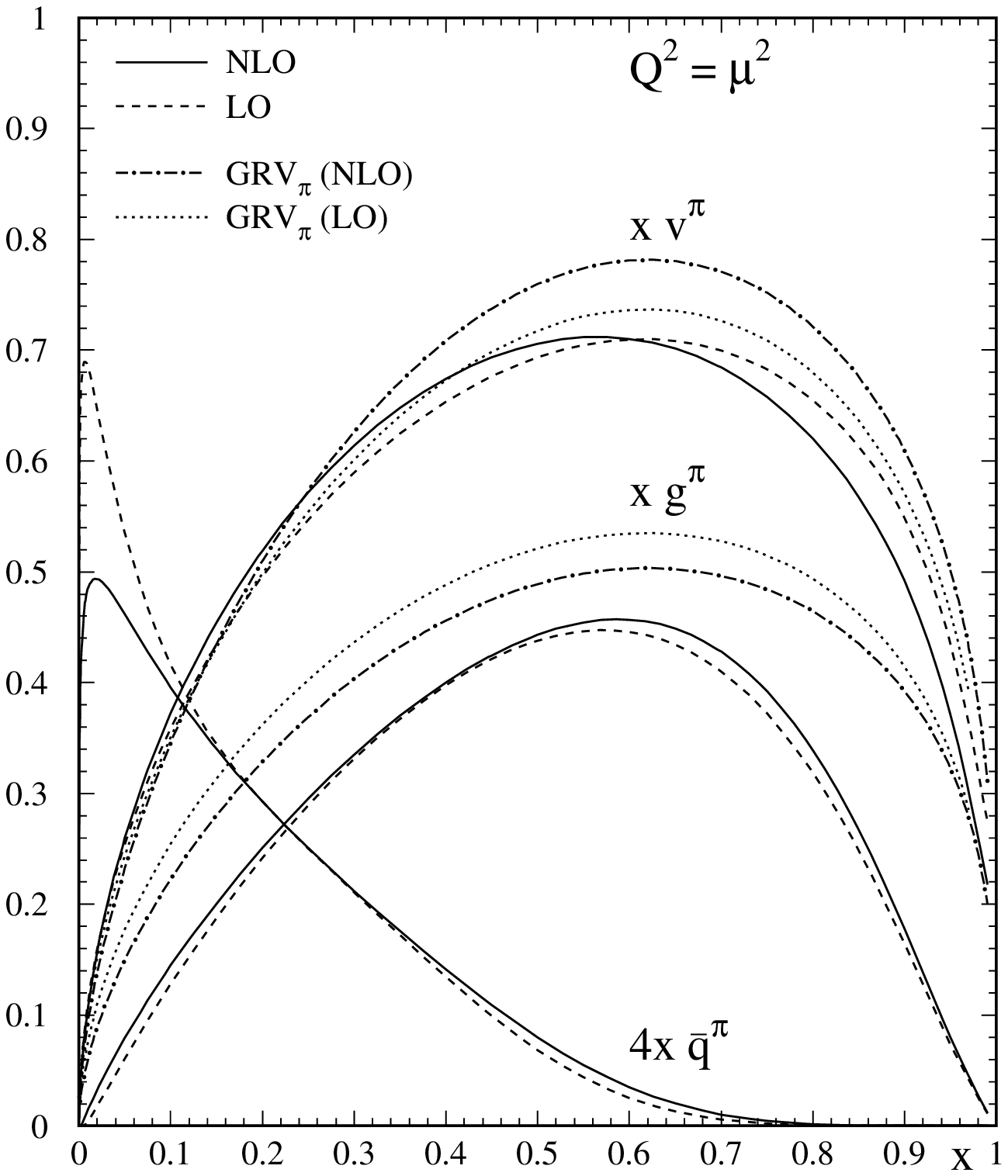,width=12cm}

\vspace*{1cm}
{\Large\bf Fig. 1}
\end{figure}
\newpage
\begin{figure}
\centering
%\vspace*{-1.5cm}
%\hspace*{-2.5cm}
\epsfig{figure=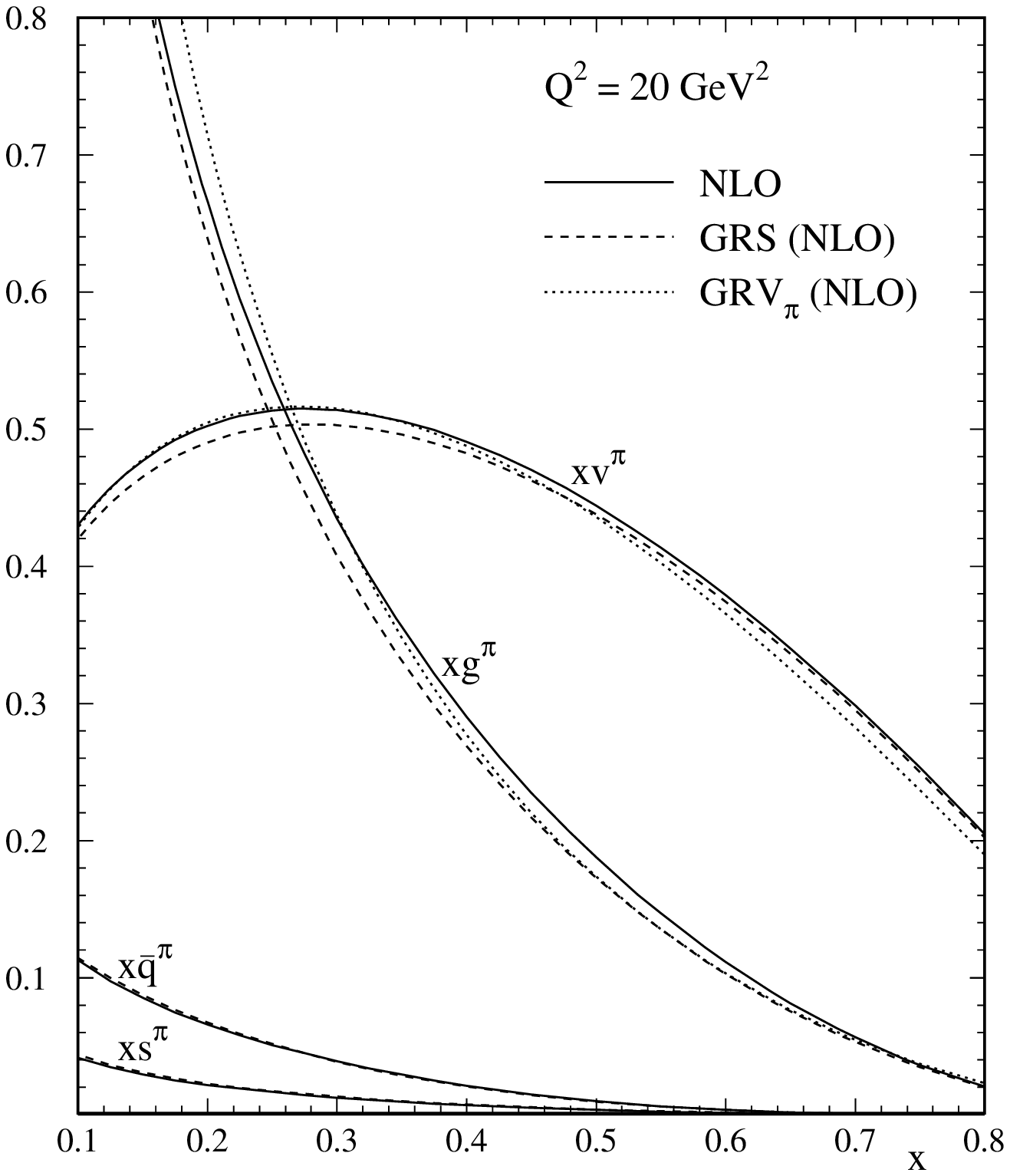,width=12cm}

\vspace*{1cm}
{\Large\bf Fig. 2}
\end{figure}
\newpage
\begin{figure}[t]
\centering
%\vspace*{-1.5cm}
%\hspace*{-2.5cm}
\epsfig{figure=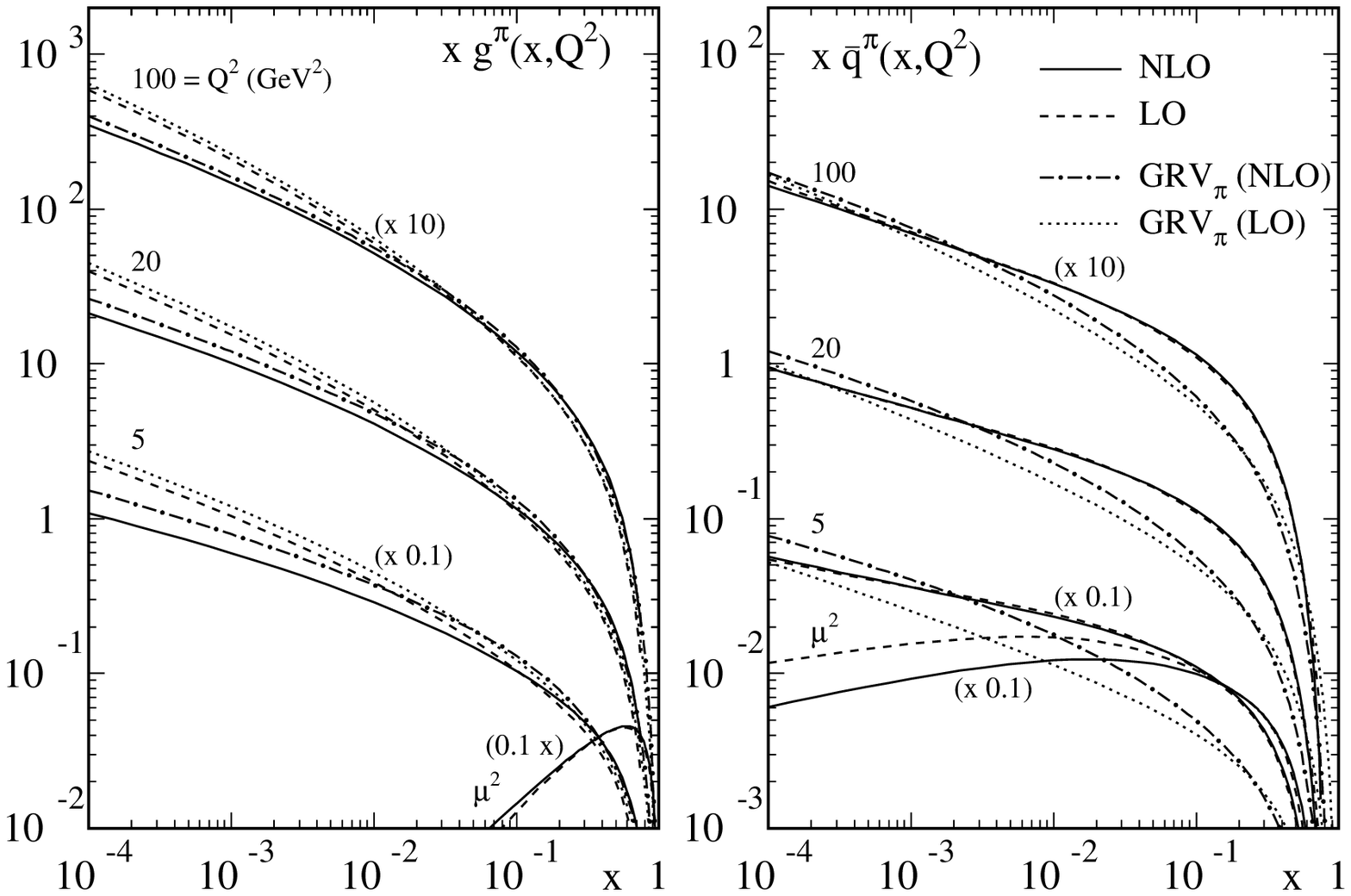,angle=90,width=12cm}

%\hspace*{-6.5cm}
\vspace*{1cm}
{\Large\bf Fig. 3}
\end{figure}
\end{document}